\documentclass[english,a4paper,twocolumn,final]{article}

\makeatletter
\def\input@path{{.}}
\makeatother

\usepackage{amsmath}
\usepackage{amssymb}
\usepackage{amsthm}
\usepackage{babel}
\usepackage{color}
\definecolor{darkblue}{cmyk}{0.83,0.89,0.0,0.43}
\usepackage[T1]{fontenc}
\usepackage{framed}
\usepackage{graphicx}
\usepackage[colorlinks,linkcolor=darkblue,urlcolor=blue,citecolor=darkblue,hyperindex,plainpages=false]{hyperref} 
\usepackage[latin1]{inputenc}
\usepackage{latex8}
\usepackage{multirow}
\usepackage{times}
\usepackage{verbatim}

\makeatletter

\newcommand{\dcsFigRef}[1]{Fig.~\ref{fig:#1}}

\newcommand{\dcsSecRef}[1]{\S\ref{sec:#1}}
\newcommand{\dcsSSecRef}[1]{\S\ref{ssec:#1}}
\newcommand{\dcsSSSecRef}[1]{\S\ref{sssec:#1}}
\newcommand{\dcsTabRef}[1]{Tab.~\ref{tbl:#1}}

\def\keywordname{{\bf Keywords:}}
\newcommand{\keywords}[1]{\par\addvspace\baselineskip
\noindent\keywordname\enspace\ignorespaces\textnormal{#1}}

\newcommand{\dcsProtoMsg}[1]{{\footnotesize\texttt{$\langle$~#1~$\rangle$}}}
\newcommand{\dcsProtoKeyword}[1]{\texttt{\textbf{#1}}}
\newcommand{\dcsProtoField}[1]{\texttt{#1}}
\newcommand{\dcsProtoFieldEnc}[1]{\ensuremath{E_{64}\left[\text{\texttt{#1}}\right]}}
\newcommand{\dcsMathId}[1]{\ensuremath{\mathit{#1}}}
\newcommand{\dcsFuncName}[1]{\ensuremath{\mathit{#1}}}

\makeatother

\begin{document}

\bibliographystyle{latex8}


\title{The TAAROA Project Specification~\thanks{This work has been supported by TOP-IX and the Piedmont Region Agency under the Innovation Development Program.}}
\author{Cosimo Anglano \and Massimo Canonico \and Marco Guazzone \and Matteo Zola\\
\centerline{Department of Computer Science, University of Piemonte Orientale, Alessandria (Italy),}\\
\centerline{email:\{cosimo.anglano,massimo.canonico,marco.guazzone,matteo.zola\}@unipmn.it}}
\date{}

\maketitle

\thispagestyle{empty}


\begin{abstract}
Since its introduction, the Grid computing paradigm has been widely adopted both in scientific and also in industrial areas.
The main advantage of the Grid computing paradigm is the ability to enable, in a transparent way, the sharing and the coordination of several heterogeneous and large-scale distributed resources belonging to different institutional domains.
One of its limitation is the lack of facilities for executing services.
In fact, Grid computing has been traditionally used and improved for running computational-intensive or data-intensive applications.
A service differs from this kind of applications in that it usually waits for requests from clients and replies with useful information; moreover, a service is typically subjected to some predefined constraints, called Service Level Agreement (SLA), including both temporal and performance restrictions.
In this paper we present the TAAROA middleware, a software system that tries to extend the traditional target of the Grid computing paradigm to include the service concept.
It attempts to accomplish its goal by using the virtualization technology.
By abstracting the hardware and software resources of a computer, virtualization brings to TAAROA two important benefits: (1) the encapsulation of the service runtime environment, and (2) the possibility, through the migration facility, to move a service from the computer where it is running to another one that hopefully reduces the risk of violating some of the SLA constraints.
In the current version of TAAROA middleware there is no explicit mechanism for achieving the level of a service as defined by the related SLA; this means that actually TAAROA is only able to provide a best-effort service.

\keywords{Grid Computing, Service Level Agreement, Virtualization.}
\end{abstract}






\section{Introduction} \label{sec:intro}

The traditional use of a distributed computing infrastructure is for executing computational-intensive or data-intensive applications for solving complex problems.
These applications are characterized by the lack of user interaction and by the high demand of computational power or storage capacity.
While almost all of the scientific applications can be considered to belong to at least one of the above categories, there are others, like the ones in the business domain, that follow different behavioral patterns.
Services are an example of such applications.

A service is an application that differs from traditional resource-intensive applications for at least two aspects: (1) there is some kind of interaction with its users, that is it spends most of its time waiting for client requests (generally issued by a user) and, upon a request arrival, replies to a requesting client with useful information, and (2) is typically subjected to predefined temporal, performance and economical constraints referred to as Service Level Agreement (SLA).
The importance of services is demonstrated by the actual trend in the development and in the deployment of applications \cite{Abrams2004SOBA,Liegl2007Strategic}: the Service Oriented Architecture (SOA) model is now a fundamental part in designing and integrating applications since it allows existing IT infrastructure and systems to achieve end-to-end enterprise connectivity by removing redundancies, generating unified collaboration tools, and streamlining IT processes \cite{Cherbakov2005Impact,Bieberstein2005Impact}.

Distributed computing paradigms still lacks of suitable mechanisms in order to execute this kind of applications.
Specifically, the two most challenging and still open problems are: (1) to decide to what physical machine assign a service for execution in order to satisfy its SLA constraints and (2) to continuously monitor the execution of the already running services for preventing and possibly reacting to SLA violations.

Finding the optimal allocation of a certain number of services to a finite number of physical machine, subjected to SLA constraints, is a computationally hard problem (NP-complete) \cite{Garey1979Computers}.
Furthermore, when no such optimal allocation can be found, a quasi-optimal allocation is still required in order to minimize SLA violations; in this case, additional issues must be taken into consideration for the conflicting nature of the problem: two or more services competes for getting assigned to the machine that allows them to meet the largest number of SLA constraints.

The other challenging problem is the monitoring and the fulfillment of SLA constraints.
It consists in observing the behavior of the service execution, collecting the performance measures related to SLA constraints, predicting future behavior (on the basis of a specific performance model) and finally properly reacting in order to prevent SLA violations, without compromising the SLA constraints of other running services.

In this paper we present the TAAROA middleware, a software system that tries to extend the traditional target of the Grid computing paradigm to include the service concept.
It attempts to accomplish its goal by using the virtualization technology.
By abstracting the hardware and software resources of a computer, virtualization brings to TAAROA two important benefits: (1) the encapsulation of the service runtime environment, and (2) the opportunity, through the migration facility, of moving an executing service from the physical machine where it is running to another one which, hopefully, reduces the risk of violating some of the SLA constraints.
Even if the ultimate goal of TAAROA middleware is to enable the optimal execution of services on distributed systems, in its current version there is no explicit mechanism for achieving the level of a service as defined by the related SLA; this means that actually TAAROA is only able to provide a best-effort service.

The rest of this paper is organized as follows.
In \dcsSecRef{overview} we aim to provide an understanding of the main ideas underlying the TAAROA middleware by outlining its high-level architecture.
In \dcsSecRef{arch} we add details to the architecture described in \dcsSecRef{overview} by illustrating the role of each TAAROA component and the associated interactions.
In \dcsSecRef{db} we describe what are the information maintained by the TAAROA middleware, to enable interaction between its component, and how they are organized.
In \dcsSecRef{protocol} the communication protocol, used by TAAROA components for interacting, is thoroughly described.
Finally, in \dcsSecRef{conclusion} we provide conclusions and our future research directions.



\section{Overview of TAAROA} \label{sec:overview}

TAAROA is a software middleware which tries to enable Grid systems (i.e., distributed systems using the Grid computing paradigm) to execute services and, at the same time, to preserve their SLA constraints.
In this section we provide an high-level overview of the architecture of the TAAROA middleware.

\begin{figure*}
\centering
\includegraphics[scale=0.60,angle=270]{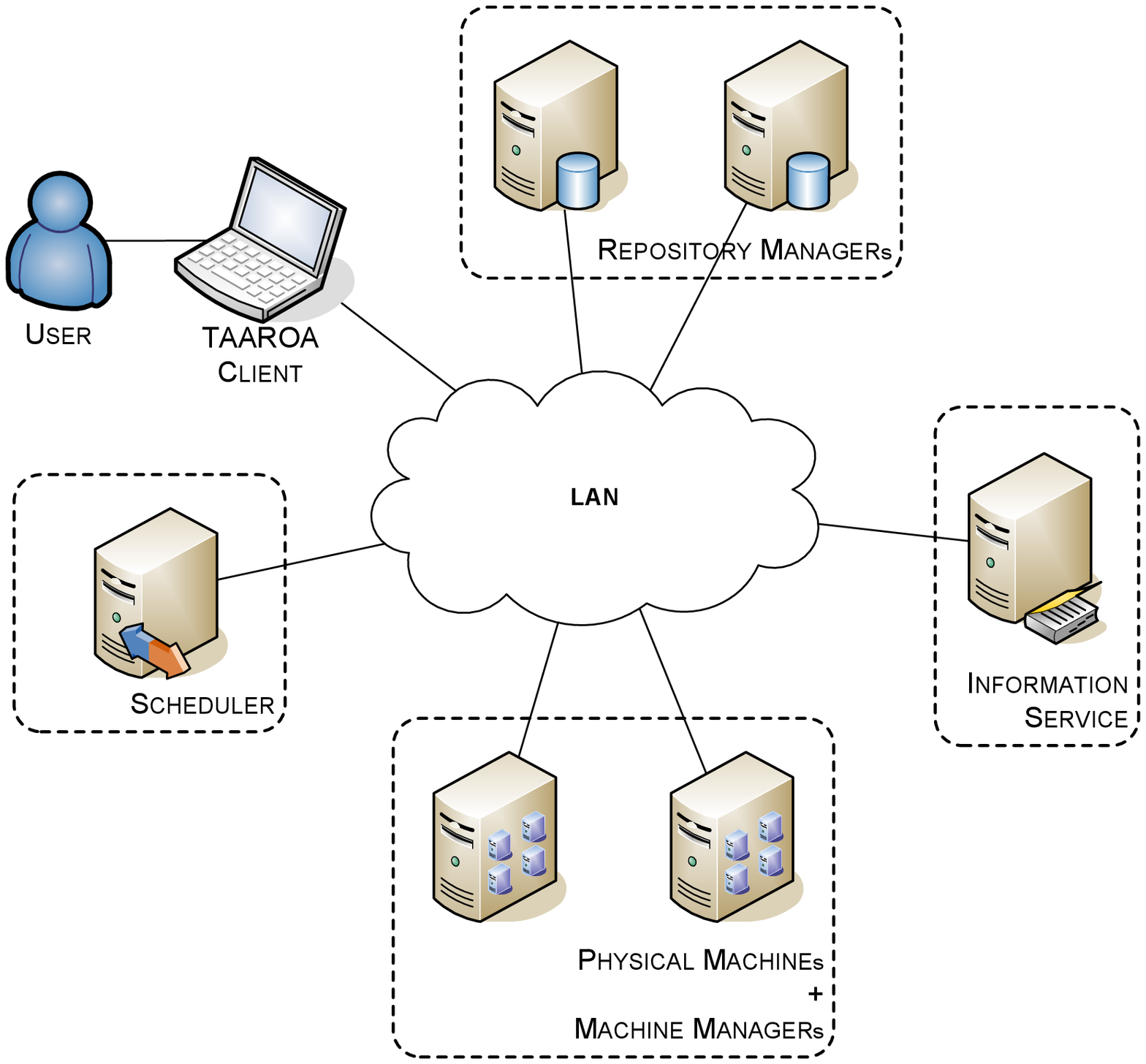}
\caption{The component-level architecture of TAAROA.}
\label{fig:arch}
\end{figure*}

As shown in \dcsFigRef{arch}, the architecture of TAAROA consists of five types of components: the TAAROA Client, the Information Service, the Repository Manager, the Scheduler and the Machine Manager (along with the associated Physical Machine).
Each component is loosely coupled to each other, meaning that components, both of the same and of different type, weakly depend from each other.

The TAAROA Client component is the user interface to the TAAROA middleware; it is the component where all user requests start.
For instance, it allows a user to execute a new service or to stop the execution of a running one.

The Information Service component is responsible for collecting, managing and publishing information about the state of the user services and of the other TAAROA components.
This information is further used by TAAROA middleware in order to locate a particular TAAROA component or to get insights about its last published state.

The Machine Manager component is in charge of executing a user service on a Physical Machine.
It uses the virtualization technology of the underlying Physical Machine in order to execute the service in an isolated environment and to avoid runtime dependencies issues: every service, along with its runtime environment, is encapsulated inside a Virtual Machine.

The Repository Manager component is responsible for hosting the images of the Virtual Machines associated to the user services.

The Scheduler component has the role of deciding to what Physical Machine a user service is to be assigned for the execution.
In the current implementation of the TAAROA middleware, requests for service submission are queued and served in the order of their arrival, by assigning to each service the first available Physical Machine.

There are several type of interactions occurring between TAAROA components.
In the following, just for illustration, we outline the workflow associated to the submission of a user service.
When a user wants to execute a service to TAAROA, he accesses to the TAAROA Client, selects the wanted service from a list retrieved by the TAAROA Client from the Information Service, and finally submit it to TAAROA.
In response to the service submission request, the TAAROA Client contacts the Scheduler for submitting the interested service.
In turn, the Scheduler queries the Information Service for obtaining the list of the available Physical Machines, along with their allocation statistics, and chooses the best one where executing the user service.
This choice can be done according to different strategies; the scheduling strategy actually implemented in TAAROA follows the FCFS (First-Come First-Served) policy both for selecting what service to execute and for choosing the Physical Machine where running it.
After having chosen the Physical Machine, the Scheduler instructs the Repository Manager to execute the service on the chosen Physical Machine.
In turn, the Repository Manager contacts the Machine Manager of the chosen Physical Machine and sends it the files related to the Virtual Machine representing the given service.
The Machine Manager, upon completion of the Virtual Machine files transfer, instructs the virtualization layer of the Physical Machine where it is running to start the new Virtual Machine.

In the above workflow, whenever a TAAROA component needs to locate another one, the Information Service is contacted in order to obtain the needed information.
The interactions occurring between the various components is describe in the following section, while the details about all of the exchanged information is delayed to \dcsSecRef{protocol}.



\section{Architecture} \label{sec:arch}

In this section we provide a detailed view of the architecture of TAAROA middleware.
We begin by a thorough description of the concepts underlying the TAAROA middleware; then we describe the role of each TAAROA components; finally we provide an example of the most important interactions between TAAROA components.
%


\subsection{Concepts} \label{ssec:concepts}

In this section we outline the fundamental concepts of the TAAROA architecture upon which all of the TAAROA components are based on.
	They include the concept of Physical Machine, Service and Virtual Machine.

\subsubsection{Physical Machine} \label{sssec:concepts-pm}

The \emph{Physical Machine} concept represents a computer connected to a network and equipped with hardware and operating system that supports virtualization technologies.

The amount of its hardware components (e.g., RAM size, disk space, CPU clock frequency, network bandwidth, and so on) is sent by the Machine Manager to the Information Service when it enters into TAAROA.
This information might be used by the Scheduler for understanding which Physical Machine best fits the performance needs of a Virtual Machine.

\subsubsection{Service} \label{sssec:concepts-svc}

The \emph{Service} concept represents an application that usually waits for incoming client requests and replies with useful information.
Typical example of Services are web servers, database servers, name servers and authentication servers.
The execution of a Service is requested through a TAAROA Client and is performed on a Physical Machine inside a Virtual Machine.
Along with the Service application, a Service usually comes with one or more predefined constraints called Service Level Agreement (SLA).
An SLA is a formal description of the level of a Service, upon which two negotiating parties (the provider and the recipient) agree.
It is commonly described in terms of Service Level Objectives (SLOs) and Service Level Specifications (SLSs), which in turn define temporal and performance metrics for measuring the level of Service, and the operational guidelines for achieving the desired level of Service, respectively.
To illustrate, an SLA may specify the levels of availability, reliability and performance of the service and possible penalties in case of violation.

In TAAROA, the application runtime environment of a Service is generally stored as one or more files; for instance, it can represent an image (e.g., an ISO-9660 file) of an operating system.
For each Service, its associated SLA is defined as a set of requirements on physical resources that needs to be satisfied once the related Service is chosen to be executed; for instance, a requirement can specify the minimum amount of space on a physical disk.

\subsubsection{Virtual Machine} \label{sssec:concepts-vm}

The \emph{Virtual Machine} concept represents a Service submitted for execution.
It is characterized by the following attributes: the instantiated Service defining a self-contained application environment, the address of the Physical Machine where the service is running, and the name or the IP address of the virtual host through which it is possible to communicate with the Virtual Machine.

\begin{figure}
\centering
\includegraphics[scale=0.42]{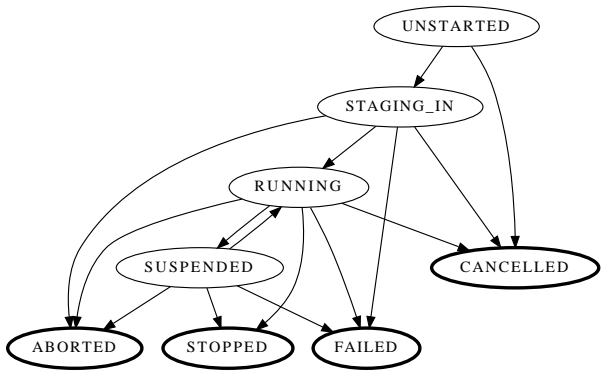}
\caption{Execution state for a Virtual Machine.}
\label{fig:vm-execstatus}
\end{figure}

The execution of a Virtual Machine can evolve through several states.
In \dcsFigRef{vm-execstatus} is shown how execution states of a Virtual Machine are related to each other.
Before running, the execution state of a Virtual Machine is UNSTARTED.
Once the Virtual Machine is selected for execution, it changes its state to STAGING\_IN; in this state, the Virtual Machine is sent to the Physical Machine where it will be executed.
After the stage-in phase has been completed, the Virtual Machine is started and its execution state becomes RUNNING.
Eventually, when the Virtual Machine is no more needed, the TAAROA Client can decide to shut it down, making the execution state of that Virtual Machine to switch to STOPPED.
There are other possible execution states a Virtual Machine can take.
Specifically, if the execution of a Virtual Machine is explicitly cancelled by a TAAROA Client, the corresponding execution state changes to CANCELLED.
If the execution of a Virtual Machine is prematurely arrested by a TAAROA component (e.g., for the lack of physical resources needed by the Virtual Machine), the corresponding execution state changes to ABORTED.
If something goes wrong during the execution of a Virtual Machine, the corresponding execution state changes to FAILED.
The execution of a Virtual Machine can even be temporarily suspended; in this case the execution state changes to SUSPENDED.
Once its execution is resumed, the execution state turns back to RUNNING.

Execution states divide in two main groups: \emph{final} and \emph{temporary} states.
Final states are the ones that once a Virtual Machine enters, cannot leave any more.
They include the ABORTED, CANCELLED, FAILED and STOPPED states (i.e., the ones marked with a thicker line in the figure).
Instead, temporary states are the ones that a Virtual Machine temporarily takes before entering a final state.
They include the RUNNING, STAGING\_IN, SUSPENDED and UNSTARTED states.
According to \dcsFigRef{vm-execstatus}, the execution of a Virtual Machine, generally, cycles through one of more temporary states before entering a final state.



\subsection{Components} \label{ssec:components}

In this section we provide a detailed description of the components of the TAAROA architecture.
As described in \dcsSecRef{overview} and shown on \dcsFigRef{arch}, there are five types of components: the TAAROA Client, the Information Service, the Repository Manager, the Scheduler and the Machine Manager (along with the associated Physical Machine).
Each component is loosely coupled to each other, meaning that components, both of the same and of different type, weakly depend from each other.
Components, in opposite to concepts, interact directly with the TAAROA system to achieve their goals.
There are \emph{active} and \emph{passive} components.
Active components are the ones that initiate interactions with the middleware; in TAAROA, the TAAROA Client, the Repository Manager and the Machine Manager are active components.
Passive components are the ones that act as a consequence of stimulus sent by active components to the system; in TAAROA, the Information Service and the Scheduler are passive components.

\subsubsection{Information Service} \label{sssec:components-is}

The \emph{Information Service} component is responsible for collecting, managing and publishing information about the primary TAAROA entities: Physical Machine, Repository Manager, Service and Virtual Machine. 
The structure of the information stored in the Information Service follows the database schema described in \dcsSecRef{db}.

Every TAAROA component can communicate with the Information Service.
The Repository Manager component contacts the Information Service for registering or unregistering itself; in addition, it is responsible for inserting, updating and removing information about Services.
The Machine Manager component interacts with the Information Service for registering or unregistering itself, the Physical Machine where it runs and the Virtual Machines hosted by the Physical Machine.
The TAAROA Client component asks the Information Service to provide it the list of Services for choosing what of them to submit for execution.
The Scheduler component queries the Information Service for obtaining the resource utilization of the available Physical Machines; this information can help it to decide on what machine a given Service is to be executed.

Actually, there is only one Information Service component in TAAROA.
This means that information about the state of TAAROA resources are definitively managed in a centralized way.

\subsubsection{Machine Manager} \label{sssec:components-mm}

The \emph{Machine Manager} component is a software component that runs on a Physical Machine and waits for connections.

The Machine Manager is in charge of managing each Virtual Machine that is hosted by the Physical Machine where it is running, by means of a software component named Virtual Machine Monitor (VMM).
The VMM, also referred to as hypervisor, is a software layer that provides virtualization support and can run either directly on hardware, if that supports it, or on top of the operating system.

At startup the Machine Manager contacts the Information Service and sends it the hardware characteristics of the Physical Machine on which resides.
When the Repository Manager is asked to start a Service, it sends to the Machine Manager the Virtual Machine image and the related configuration file.
The Virtual Machine image contains all the files that build up the Service runtime environment, while the configuration file contains all the settings used by the VMM when starting, stopping and suspending the Virtual Machine.

A Machine Manager is also responsible for keeping the Information Service up to date by communicating changes in the amount of resources allocated to every managed Virtual Machine.

\subsubsection{Repository Manager} \label{sssec:components-rm}

The \emph{Repository Manager} component is a software component that is responsible for hosting Virtual Machine images.
It has access to a storage area (either local or remote) where each Virtual Machine image is placed, along with its configuration file ad everything that is necessary for running the Virtual Machine.

Upon service submission request, it takes care of sending the corresponding Virtual Machine image and related files, to a given Physical Machine and asks the Machine Manager (located on that machine) to start the Virtual Machine.
In a similar way, when it is asked to stop a Virtual Machine, it instructs the Machine Manager (where the Virtual Machine is running) accordingly.

A Repository Manager is also responsible for keeping the Information Service up to date by communicating changes in the execution state of every managed Virtual Machine.

\subsubsection{Scheduler} \label{sssec:components-sc}

The \emph{Scheduler} component is responsible for deciding to what Physical Machine a Service is to be assigned for the execution.
This decision is taken through the so called scheduling policy, often referred to as scheduling heuristic.
In the current implementation of the TAAROA middleware, the only available scheduling heuristic is the one that uses the First-Come-First-Served (FCFS) policy: requests for Service submission are queued and served in the order of their arrival, by assigning to each Service the first available Physical Machine.

Actually, there is only one Scheduler component in TAAROA.
This means that, from the point of view of a TAAROA Client, the Scheduler component is the only single point of control for starting, monitoring and stopping the execution of a Service in TAAROA: every request for submitting, managing and terminating a Service must go through the Scheduler component.

\subsubsection{TAAROA Client} \label{sssec:components-tc}

The \emph{TAAROA Client} component communicates with the other TAAROA components in order to submit Services and to monitor their execution.
It represents the user interface to the TAAROA middleware: all of the interactions occurring between a user (or a user application) and TAAROA happen through this component.

The \emph{TAAROA Web Portal} is a specific type of TAAROA Client which offers an high-level user-friendly interface in order to make the interactions with the other TAAROA components easier from the point of view of its users.



\subsection{Workflow} \label{ssec:workflow}

In this section we describe the static relations and the dynamic interactions between the TAAROA components.

\begin{figure}
\centering
\centerline{\includegraphics[scale=0.55]{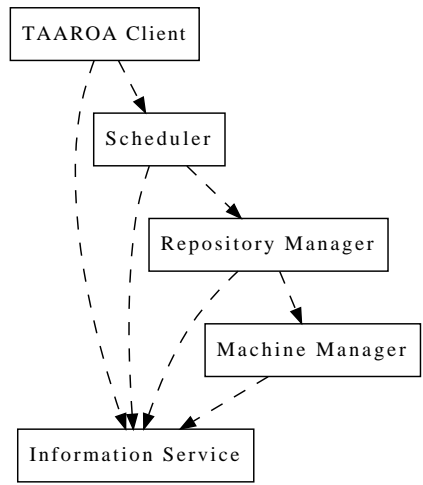}}
\caption{Static relations between TAAROA components.}
\label{fig:workflow-relations}
\end{figure}

\dcsFigRef{workflow-relations} shows the structural relations between TAAROA components.
The TAAROA Client uses the Information Service for getting access to information about Services and other related entities; for instance, it queries the Information Service for retrieving the list of available Services.
In addition, the TAAROA Client uses the Scheduler for executing, managing and stopping one or more Services.
The Scheduler uses the Information Service for obtaining information about Services and resource utilization of Physical Machines.
This information might be used, for example, for deciding to what Physical Machine a given Service is to be assigned for the execution.
Moreover, the Scheduler uses the Repository Manager for submitting and stopping a specific Service.
The Repository Manager uses the Information Service for registering and unregistering itself along with the published Services.
The Repository Manager also uses the Machine Manager for controlling the execution of a Virtual Machine.
Finally, the Machine Manager uses the Information Service for registering and unregistering the Physical Machine on which it runs and every Virtual Machine it manages.

The dynamic interactions between TAAROA components are based on a message-oriented and stateless protocol; this means that each pair of TAAROA components exchanges messages with the rest of TAAROA components, in the form of request-reply messages, and each message neither depends on previously sent messages nor on additional information stored on the receiving component.

In the rest of this section we present the workflow concerning two of the most important TAAROA interactions: the Service submission workflow and the Service stopping workflow.
The type of modelling diagram used for describing these interactions is the UML \emph{communication diagram} \cite{OMG2007UML}.
This kind of diagram models the interactions between TAAROA components by showing the flow of messages exchanged among them.
In order to maintain the sequential ordering of interactions, messages are labeled with a chronological number (usually starting from one), placed near the link where the message is sent over.

\begin{figure*}
\centering
\centerline{\includegraphics[scale=0.54,angle=270]{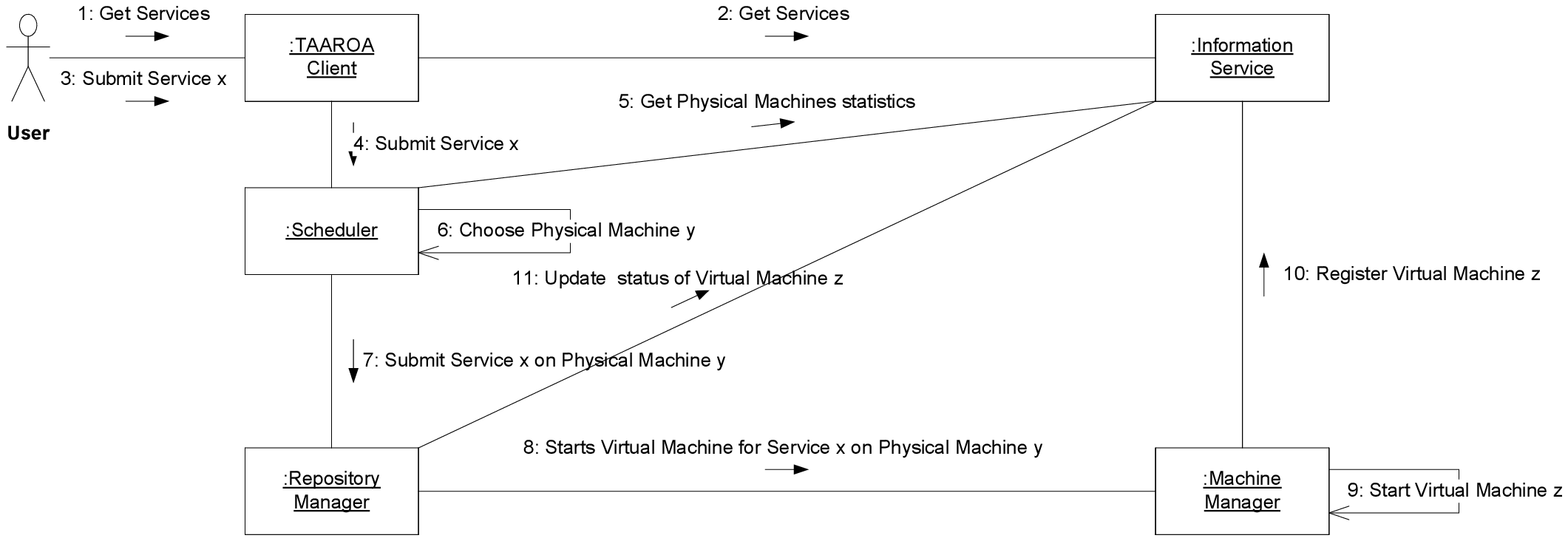}}
\caption{The communication diagram for TAAROA service submission.}
\label{fig:workflow-svcstart-col}
\end{figure*}

In the Service submission workflow, schematically shown on \dcsFigRef{workflow-svcstart-col}, the active actor is the TAAROA Client.
When it wants to submit a Service, the first action it performs is to request to the Information Service component the list of the available Services.
After having chosen the Service it wants to run, denoted with Service $x$ in the figure, it sends a request to the Scheduler component for submitting the selected Service and waits for a replay.
In turn, the Scheduler contacts the Information Service for obtaining the list of Physical Machines along with their allocation statistics.
The information needed to calculate these statistics is sent to the Information Service by the Machine Manager both when the Physical Machine, on which it is running, is added to TAAROA and when one of the Virtual Machine, running on that Physical Machine, changes its execution state.
These statistics can be used by the Scheduler for deciding, according to a proper scheduling heuristic, if a Service can be immediately executed and on what Physical Machine.
If the Service can be executed and a suitable Physical Machine is found, the Scheduler instructs the Repository Manager to submit the Virtual Machine associated to this Service on the chosen Physical Machine.
In the figure, the Physical Machine chosen by the Scheduler is named $y$.
Then, the Repository Manager contacts the Machine Manager of the chosen Physical Machine for starting a new Virtual Machine for the given Service.
It sends all the files composing the Virtual Machine to the Machine Manager which, in turn, instructs the virtualization layer to start the Virtual Machine.
In the figure, the new Virtual Machine is marked as $z$.
In case of success, it registers the newly created Virtual Machine to the Information Service along with its parameters (network configuration, administration credentials and so on), for later retrieval, and notifies the Repository Manager about the starting of the Virtual Machine.
When the Repository Manager receives the notification about the execution of the Virtual Machine, it contacts the Information Service, for updating the execution status of that Virtual Machine, and the Scheduler, for passing to it the Virtual Machine global identifier (i.e., the value that uniquely identifies the Virtual Machine inside the TAAROA system) obtained from the Information Service.
The Scheduler, in turn, notifies the requesting TAAROA Client, that is the one that initially began the workflow.
In the above figure, all the replies are omitted for the sake of simplicity.

\begin{figure*}
\centering
\includegraphics[scale=0.54,angle=270]{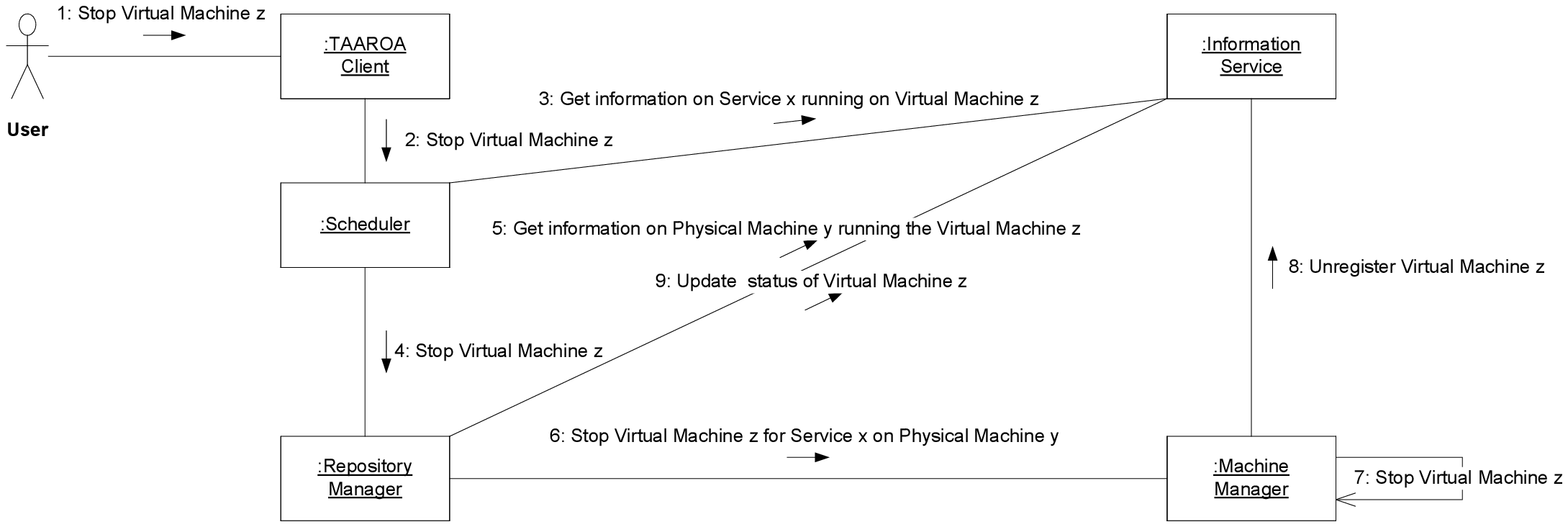}
\caption{The communication diagram for TAAROA service stopping.}
\label{fig:workflow-svcstop-col}
\end{figure*}

The other important interaction between TAAROA components is the stopping of a Service.
In \dcsFigRef{workflow-svcstop-col} is depicted the workflow for stopping a Service.
Likewise the Service submission workflow, the active actor is the TAAROA Client.
When it wants to stop a running Service, identified as $z$ in the figure, it sends a request to the Scheduler component.
Subsequently, the Scheduler contacts the Repository Manager component which, in turn, queries the Information Service for finding out what Physical Machine is hosting that Virtual Machine.
Once the Repository Manager obtains the requested information, it asks the Machine Manager, running on that Physical Machine, to stop the involved Virtual Machine.
In the figure, the hosting Physical Machine is denoted with $y$.
The Machine Manager delegates the Virtual Machine Monitor to stop that Virtual Machine, then unregisters the Virtual Machine on the Information Service and, on success, notifies the Repository Manager about the stopping of the Virtual Machine.
Finally, the Repository Manager updates the execution status of the stopped Virtual Machine and then notifies the Scheduler, which successively notifies the requesting TAAROA Client.
Similarly to the diagram for the Service submission workflow, in the above figure, all the replies are omitted for the sake of simplicity.

The other kind of interactions that might occur between TAAROA components mainly concern the updating of information kept in the Information Service.
For example, each Repository Manager registers or unregisters itself to the Information Service whenever it joins to or leaves the TAAROA system, respectively.
Likewise, when a Machine Manager joins to or leaves the TAAROA system, it registers or unregisters itself, respectively, to the Information Service component, along with the Physical Machine on which it runs.
Moreover, every time a new Service is added or an existing Service is removed from TAAROA, it is stored on or deleted from a Repository Manager which, in turn, takes care of registering or unregistering it to the Information Service, respectively.
When a Virtual Machine changes the state of its execution, the Repository Manager updates the related information maintained by the Information Service accordingly.




\section{Database} \label{sec:db}

In this section we describe the databases maintained by the Information Service, the Repository Manager and the Machine Manager components.
The modelling diagram used for showing the different database models is the Data Structure Diagram (DSD).
This kind of diagram is an extension of the classic Entity-Relationship (E-R) diagram \cite{Chen1976ER}; it differs from it in that the E-R model focuses on the relationships between different entities, whereas a DSD focuses on the relationships of the elements within an entity, enabling users to better understand the links and the relationships between each entity.
In this diagram, entities are represented as boxes, entity attributes are specified inside the entity boxes, while binary relationships are drawn as lines connecting the boxes representing the participating entities.
For $n$-ary relationships, an additional entity is used; it might have attributes which specify the constraints that bind participating entities together.
The cardinality of an entity for a particular relationship is expressed using the ``crow's foot'' notation.

\subsection{The \emph{Information Service} database} \label{ssec:db-is}

\begin{figure*}
\centering
\includegraphics[scale=0.60]{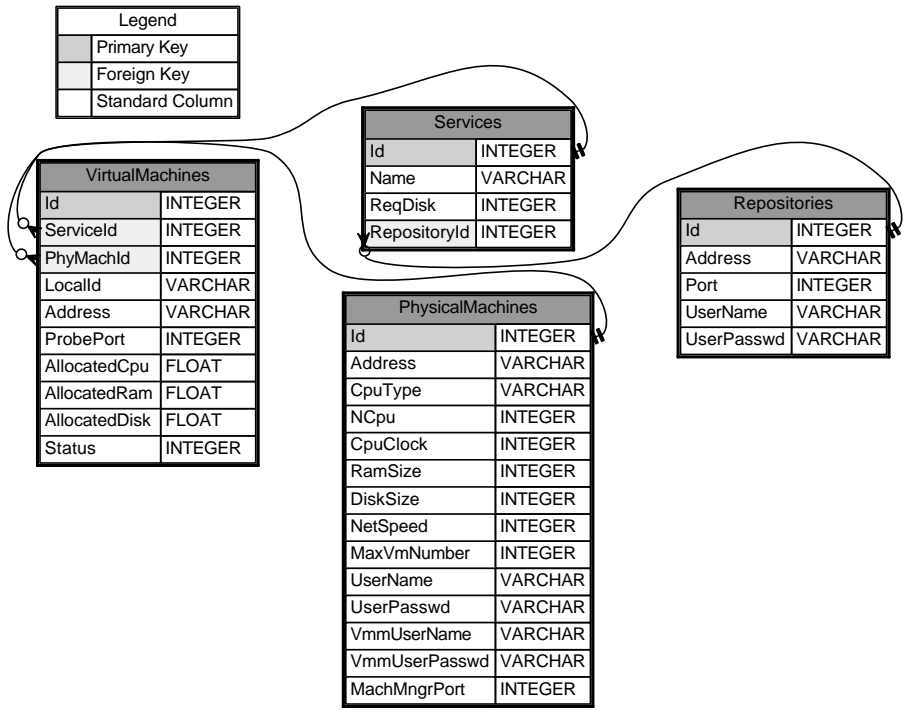}
\caption{The E-R schema of the Information Service database.}
\label{fig:db-is-schema}
\end{figure*}

In \dcsFigRef{db-is-schema} is shown the DSD of the database used by the Information Service component.
In the rest of this section, we describe the entities and the relationships contained in this database.

\subsubsection{The \emph{PhysicalMachines} entity} \label{sssec:db-is-pm}

The \emph{PhysicalMachines} entity represents the Physical Machine concept presented in \dcsSSSecRef{concepts-pm}.
A PhysicalMachines entity is uniquely identified by an integer positive number represented by the attribute \emph{Id}.
A Physical Machine can host one or more Virtual Machines, whereas a Virtual Machine is running on only one Physical Machine at a time.
The maximum amount of Virtual Machines a given Physical Machine can run is specified by the attribute \emph{MaxVmNumber}.

The other attributes concern hardware and system properties and administration information.
Regarding the hardware and system characteristics, the attribute \emph{CpuType} represents the vendor and the model of the CPU, the attribute \emph{NCpu} denotes the number of cores or processors installed on the Physical Machine, while the attribute \emph{CpuClock} specifies the CPU clock frequency (in MegaHertz).
The attributes \emph{RamSize}, \emph{DiskSize} and \emph{NetSpeed} represent respectively the amount of system RAM (in MegaBytes), the amount of disk space (in MegaBytes) and the speed of the network card (in MegaBps) of the Physical Machine.
Finally, the attribute \emph{Address} represents the IP address of the Physical Machine.

For what concerns the information for administration purpose, the attributes \emph{UserName} and \emph{UserPassword} represent the credentials for remotely accessing to the Physical Machine, the attribute \emph{VmmUserName} and \emph{VmmUserPassword} are the credentials for gaining access to the Virtual Machine Monitor and the attribute \emph{MachMngrPort} is the port at which the Machine Manager waits for requests.

\subsubsection{The \emph{Repositories} entity} \label{sssec:db-is-rm}

The \emph{Repositories} entity represents the Repository Manager concept stated in \dcsSSSecRef{components-rm}.
A Repositories entity is uniquely identified by the attribute \emph{Id}, which is an integer positive number.
The attributes \emph{Address} and \emph{Port} are used for connecting to the Repository Manager, while attributes \emph{UserName} and \emph{UserPasswd} are the credentials needed for gaining access to it.

\subsubsection{The \emph{Services} entity} \label{sssec:db-is-svc}

The \emph{Services} entity models the Service concept described in \dcsSSSecRef{concepts-svc}.
A Services entity is uniquely identified by an integer positive number represented by the attribute \emph{Id}.
This entity is uniquely associated to a Repositories entity (through the attribute \emph{RepositoryId}), meaning that a Service is provided by one and only one Repository Manager component.
Furthermore, a Services entity can participate in the association with one more VirtualMachines entities, but a VirtualMachines entity is associated to exactly one Services entity.
This basically means that the same Service can appear in one or more Virtual Machines, but a Virtual Machine can only run exactly one Service.
The remaining attributes are the Service name (attribute \emph{Name}) and the amount of disk space (in bytes) needed by the Service for executing (attribute \emph{ReqDisk}).

\subsubsection{The \emph{VirtualMachines} entity} \label{sssec:db-is-vm}

The \emph{VirtualMachines} entity describes the Virtual Machine concept outlined in \dcsSSSecRef{concepts-vm}.
A VirtualMachines entity is uniquely identified by the attribute \emph{Id}, which is an integer positive number.
A Virtual Machine represents a solely running Service instance and can live in only one Physical Machine (though Virtual Machine migration can change the hosting machine along the time); the attributes that link a Virtual Machine to its Service and to its Physical Machine are \emph{ServiceId} and \emph{PhyMachId}, respectively.
On the other hand, more than one Virtual Machine can execute the same Service and a Physical Machine may contain several Virtual Machines.
Among the other attributes characterizing this entity, those that are worth noting are the ones indicating the amount of physical resources allocated to a Virtual Machine and the state of the execution.
The resource allocation attributes include the CPU share allocation (attribute \emph{AllocatedCpu}), the fraction of allocated system memory (attribute \emph{AllocatedRam}) and the fraction of allocated disk space (attribute \emph{AllocatedDisk}).
For what concerns the execution state of a Virtual Machine, it is represented by the attribute \emph{Status}, an integer number whose possible values are defined according to the TAAROA communication protocol (described in \dcsSecRef{protocol}).

\subsection{The \emph{Repository Manager} database} \label{ssec:db-rm}

\begin{figure}
\centering
\includegraphics[scale=0.70]{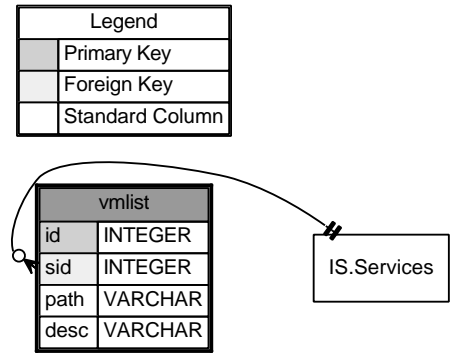}
\caption{The E-R schema of the Repository Manager database.}
\label{fig:db-rm-schema}
\end{figure}

In \dcsFigRef{db-rm-schema} is shown the DSD of the database used by the Repository Manager component.
The purpose of this database is to store information that allow to associate TAAROA global descriptors with information that are local to each Repository Manager.
For this reason, each Repository Manager maintains a different copy of this database.

In the rest of this section, we describe the entities and the relationships contained in this database.

\subsubsection{The \emph{vmlist} entity} \label{sssec:db-rm-vm}

The \emph{vmlist} entity is used by the Repository Manager for retrieving, from a given Service identifier, all of the files composing a Virtual Machine.
The attribute \emph{sid} represents the Service identifier related to a particular Virtual Machine; it is a foreign key referring to the attribute \emph{Id} of the entity \emph{Services}, stored in the Information Service database (see \dcsSSSecRef{db-is-svc}).
The \emph{path} attribute is the actual path where all the files for a given Virtual Machine are stored.

\subsection{The \emph{Machine Manager} database} \label{ssec:db-mm}

\begin{figure}
\centering
\includegraphics[scale=0.70]{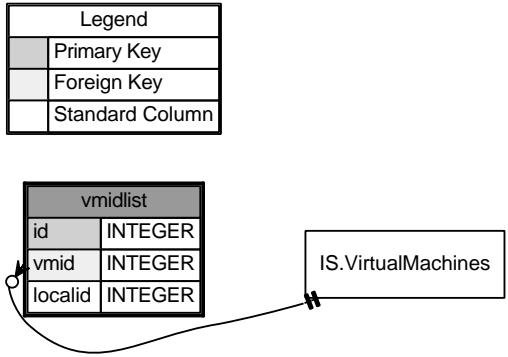}
\caption{The E-R schema of the Machine Manager database.}
\label{fig:db-mm-schema}
\end{figure}

In \dcsFigRef{db-mm-schema} is shown the DSD of the database used by the Machine Manager component.
This database contains only information that is local to each Machine Manager; for instance, the information regarding Virtual Machines is restricted only to the ones running on the Physical Machine on which the Machine Manager resides.
For this reason, each Machine Manager maintains a different copy of this database.

In the rest of this section, we describe the entities and the relationships contained in this database.

\subsubsection{The \emph{vmidlist} entity} \label{sssec:db-mm-vm}

The \emph{vmidlist} entity is used by the Machine Manager for associating a Virtual Machine concept (see \dcsSSSecRef{concepts-vm}) with a real Virtual Machine implementation.
Specifically, it links a Virtual Machine global identifier, assigned by the Information Service, to a local identifier associated to the corresponding Virtual Machine running on the Physical Machine on which the Machine Manager resides; this local identifier is assigned to the real Virtual Machine by the underlying Virtual Machine Monitor.
Each vmidlist entity is uniquely identified by an integer positive number represented by the attribute \emph{Id}.
The attribute \emph{vmid} represents the TAAROA Virtual Machine identifier, whereas the attribute \emph{localid} is the local Virtual Machine identifier assigned by the Virtual Machine Monitor.



\section{Communication Protocol} \label{sec:protocol}

In this section we describe the communication protocol used by the current version of TAAROA middleware.
The protocol is at the base of all the dynamic interactions occurring between TAAROA components; it is a message-oriented and stateless protocol, that is each TAAROA component exchanges with the others a series of request-reply messages that neither depend on previously sent messages nor on additional information stored on the receiving component.

The specification of the protocol messages follows precise symbol and number conventions.
In \dcsTabRef{notations} are shown the symbols, along with their meaning, employed for describing the format of the protocol messages.
Instead, in \dcsSSecRef{formats}, the format of the number, the unit of measurement and the other constants is illustrated.



\begin{table}
\center
{\small
\begin{tabular}{ll}
\hline\hline
Symbol & Description \tabularnewline
\hline
<\#> & The literal character '\texttt{\#}'. \tabularnewline
<b> & A sequence of one or more blank characters \tabularnewline
    & (whitespace or horizontal tabulation). \tabularnewline
$D_{64}[s]$ & The base64 decoding of the string $s$. \tabularnewline
$E_{64}[s]$ & The base64 encoding of the string $s$. \tabularnewline
IS & Abbreviation for Information Server. \tabularnewline
MM & Abbreviation for Machine Manager. \tabularnewline
RM & Abbreviation for Repository Manager. \tabularnewline
SC & Abbreviation for Scheduler. \tabularnewline
SVC & Abbreviation for Service. \tabularnewline
TC & Abbreviation for TAAROA client. \tabularnewline
VM & Abbreviation for Virtual Machine. \tabularnewline
WP & Abbreviation for TAAROA Web Portal. \tabularnewline
\hline\hline
\end{tabular}
\caption{Notations and abbreviations for the TAAROA communication protocol.}
\label{tbl:notations}
}
\end{table}



\subsection{Common formats} \label{ssec:formats}

\subsubsection{Integer Numbers representation} \label{sssec:formats-intnum}

The protocol supports the following integer number format (expressed as POSIX regular expression):
\begin{itemize}
\item $\textbackslash d+$ (e.g. $14$).
\end{itemize}
No negative value is allowed.

\subsubsection{Real Numbers representation} \label{sssec:formats-realnum}

The protocol supports the following real number formats (expressed as POSIX regular expression):
\begin{itemize}
\item Standard notation: $\textbackslash d+ \textbackslash \ldotp \textbackslash d+$ (e.g. $14.5$).
\item Scientific notation: $\textbackslash d+ \textbackslash \ldotp \textbackslash d+[eE][+-] \textbackslash d+$ (e.g. $1.45e+1$).
\end{itemize}
No negative value is allowed.

\subsubsection{Frequency Unit of Measurement representation} \label{sssec:formats-freq}

The string MUST be an integer number optionally followed by a unit specifier character.
Possible unit specifiers are:
\begin{itemize}
\item \dcsProtoField{Hz} for Hertz.
\item \dcsProtoField{KHz} for KiloHertz.
\item \dcsProtoField{MHz} for MegaHertz.
\item \dcsProtoField{GHz} for GigaHertz.
\item \dcsProtoField{THz} for TeraHertz.
\item \dcsProtoField{PHz} for PetaHertz.
\end{itemize}
If no unit specifier character is specified, the default value depends on the context where the unit of measurement has to be specified.

\subsubsection{Memory Unit of Measurement representation} \label{sssec:formats-mem}

The string MUST be an integer number optionally followed by a unit specifier character.
Possible unit specifiers are:
\begin{itemize}
\item \dcsProtoField{B} for bytes.
\item \dcsProtoField{KB} for Kilobytes.
\item \dcsProtoField{MB} for Megabytes.
\item \dcsProtoField{GB} for Gigabytes.
\item \dcsProtoField{TB} for Terabytes.
\item \dcsProtoField{PB} for Petabytes.
\end{itemize}
If no unit specifier character is specified, the default value depends on the context where the unit of measurement has to be specified.

\subsubsection{Net Speed Unit of Measurement representation} \label{sssec:formats-net}

The string MUST be an integer number \dcsSSSecRef{formats-intnum} optionally followed by a unit specifier character.
Possible unit specifiers are:
\begin{itemize}
\item \dcsProtoField{bps} for bits-per-second (bit/s).
\item \dcsProtoField{Kbps} for Kilobps (Kbit/s).
\item \dcsProtoField{Mbps} for Megabps (Mbit/s).
\item \dcsProtoField{Gbps} for Gigabps (Gbit/s).
\item \dcsProtoField{Tbps} for Terabps (Tbit/s).
\item \dcsProtoField{Pbps} for Petabps (Pbit/s).
\end{itemize}
If no unit specifier character is specified, the default value depends on the context where the unit of measurement has to be specified.

\subsubsection{Execution Status Codification} \label{sssec:formats-execstatus}

The execution status of a Virtual Machine is coded as an integer number:
\begin{description}
\item{\dcsProtoField{0}}: represents the UNKNOWN execution status.
\item{\dcsProtoField{1}}: represents the UNSTARTED execution status.
\item{\dcsProtoField{2}}: represents the READY execution status.
\item{\dcsProtoField{3}}: represents the STAGING\_IN execution status.
\item{\dcsProtoField{4}}: represents the RUNNING execution status.
\item{\dcsProtoField{5}}: represents the SUSPENDED execution status.
\item{\dcsProtoField{6}}: represents the STOPPED execution status.
\item{\dcsProtoField{7}}: represents the CANCELLED execution status.
\item{\dcsProtoField{8}}: represents the FAILED execution status.
\item{\dcsProtoField{9}}: represents the ABORTED execution status.
\end{description}



\subsection{Messages issued to the Information Server} \label{ssec:is}

\subsubsection{GETPHYMACH -- \emph{Physical Machine details} request}

Sent by a RM to the IS for getting information about a specific physical machine.
\begin{quote}
\dcsProtoMsg{\dcsProtoKeyword{GETPHYMACH}<b>PHY\_ID}
\end{quote}
where:
\begin{itemize}
\item \dcsProtoField{PHY\_ID}: integer number \dcsSSSecRef{formats-intnum} representing the physical machine identifier.
\end{itemize}
Possible replies from the IS are:
\begin{itemize}
\item In case of success:
\begin{quote}
\dcsProtoMsg{\dcsProtoKeyword{OK}<b>PHY\_IP<b>MM\_PORT}
\end{quote}
where:
\begin{itemize}
\item \dcsProtoField{PHY\_IP}: string containing the IP address of the requested physical machine.
\item \dcsProtoField{MM\_PORT}: integer number \dcsSSSecRef{formats-intnum} representing the TCP port of the MM.
\end{itemize}
\item \dcsProtoMsg{\dcsProtoKeyword{ERR}<b>CODE} otherwise, where \dcsProtoField{CODE} is an integer number representing an error code.
\end{itemize}

\subsubsection{GETVM -- \emph{Virtual Machine details} request}

Sent by a TC to the IS when it wants to know the details regarding a given submitted service (virtual machine).
\begin{quote}
\dcsProtoMsg{\dcsProtoKeyword{GETVM}<b>VM\_ID}
\end{quote}
where:
\begin{itemize}
\item \dcsProtoField{VM\_ID}: integer number \dcsSSSecRef{formats-intnum} containing the submitted service (virtual machine) identifier.
\end{itemize}
Possible replies from the IS are:
\begin{itemize}
\item In case of success:
\newline
\begin{quote}
\dcsProtoMsg{\dcsProtoKeyword{OK}<b>S\_ID<b>PHY\_ID<b>\\VM\_LOCAL\_ID<b>VIRT\_IP<b>STATUS}
\end{quote}
where:
\begin{itemize}
\item \dcsProtoField{S\_ID}: integer number \dcsSSSecRef{formats-intnum} containing the service identifier.
\item \dcsProtoField{PHY\_ID}: integer number \dcsSSSecRef{formats-intnum} containing the identifier of the physical machine.
\item \dcsProtoField{VM\_LOCAL\_ID}: string containing the identifier used by the MM to uniquely retrieve a VM.
\item \dcsProtoField{VIRT\_IP}: string containing the IP address of the virtual machine on which the service is running.
\item \dcsProtoField{STATUS}: integer number \dcsSSSecRef{formats-intnum} representing the execution status of the submitted service (\dcsSSSecRef{concepts-vm}).
\end{itemize}
\item \dcsProtoMsg{\dcsProtoKeyword{ERR}<b>CODE} otherwise, where \dcsProtoField{CODE} is an integer number representing an error code.
\end{itemize}

\subsubsection{GETVMMACHMNGR -- \emph{Virtual Machine Machine Manager} request}

Sent by a RM (or other clients) to the IS when it wants to know the machine manager associated to a given submitted service (virtual machine).
\begin{quote}
\dcsProtoMsg{\dcsProtoKeyword{GETVMMACHMNGR}<b>VM\_ID}
\end{quote}
where:
\begin{itemize}
\item \dcsProtoField{VM\_ID}: integer number \dcsSSSecRef{formats-intnum} containing the submitted service (virtual machine) identifier.
\end{itemize}
Possible replies from the IS are:
\begin{itemize}
\item In case of success:
\begin{quote}
\dcsProtoMsg{\dcsProtoKeyword{OK}<b>PHY\_ID<b>PHY\_IP<b>\\MM\_PORT<b>VM\_LOCAL\_ID}
\end{quote}
where:
\begin{itemize}
\item \dcsProtoField{PHY\_ID}: integer number \dcsSSSecRef{formats-intnum} representing the identifier of the Physical Machine where the MM is running.
\item \dcsProtoField{PHY\_IP}: string representing the IP address of the Physical Machine where the MM is running.
\item \dcsProtoField{MM\_PORT}: integer number \dcsSSSecRef{formats-intnum} representing the TCP port of the MM.
\item \dcsProtoField{VM\_LOCAL\_ID}: string containing the identifier used by the MM to uniquely retrieve a VM.
\end{itemize}
\item \dcsProtoMsg{\dcsProtoKeyword{ERR}<b>CODE} otherwise, where \dcsProtoField{CODE} is an integer number representing an error code.
\end{itemize}

\subsubsection{GETVMSERV -- \emph{Virtual Machine Service} request}

Sent by a TC to the IS when it wants to know the service associated to a given submitted service (virtual machine).
\begin{quote}
\dcsProtoMsg{\dcsProtoKeyword{GETVMSERV}<b>VM\_ID}
\end{quote}
where:
\begin{itemize}
\item \dcsProtoField{VM\_ID}: integer number \dcsSSSecRef{formats-intnum} containing the submitted service (virtual machine) identifier.
\end{itemize}
Possible replies from the IS are:
\begin{itemize}
\item In case of success:
\begin{quote}
\dcsProtoMsg{\dcsProtoKeyword{OK}<b>S\_ID<b>\dcsProtoFieldEnc{NAME}<b>\\RM\_ID<b>RM\_IP<b>RM\_PORT}
\end{quote}
where:
\begin{itemize}
\item \dcsProtoField{S\_ID}: integer number \dcsSSSecRef{formats-intnum} representing the service identifier.
\item \dcsProtoField{NAME}: string representing the symbolic name of the service.
\item \dcsProtoField{RM\_ID}: integer number \dcsSSSecRef{formats-intnum} representing the RM identifier.
\item \dcsProtoField{RM\_IP}: string representing the IP address of the RM.
\item \dcsProtoField{RM\_PORT}: integer number \dcsSSSecRef{formats-intnum} representing the TCP port of the RM.
\end{itemize}
\item \dcsProtoMsg{\dcsProtoKeyword{ERR}<b>CODE} otherwise, where \dcsProtoField{CODE} is an integer number representing an error code.
\end{itemize}

\subsubsection{GETVMSTATUS -- \emph{Virtual Machine Status} request}

Sent by a TC to the IS when it wants to know the execution status of a given submitted service (virtual machine).
\begin{quote}
\dcsProtoMsg{\dcsProtoKeyword{GETVMSTATUS}<b>VM\_ID}
\end{quote}
where:
\begin{itemize}
\item \dcsProtoField{VM\_ID}: integer number \dcsSSSecRef{formats-intnum} containing the submitted service (virtual machine) identifier.
\end{itemize}
Possible replies from the IS are:
\begin{itemize}
\item In case of success:
\begin{quote}
\dcsProtoMsg{\dcsProtoKeyword{OK}<b>STATUS}
\end{quote}
where \dcsProtoField{STATUS} is an integer number \dcsSSSecRef{formats-intnum} representing the service execution status (\dcsSSSecRef{concepts-vm}).
\item \dcsProtoMsg{\dcsProtoKeyword{ERR}<b>CODE} otherwise, where \dcsProtoField{CODE} is an integer number representing an error code.
\end{itemize}

\subsubsection{LISTPHYMACH -- \emph{List of Physical Machines} request}

Sent by a TC to the IS when it wants to retrieve the list of all registered physical machines.
\begin{quote}
\dcsProtoMsg{\dcsProtoKeyword{LISTPHYMACH}}
\end{quote}
Possible replies from the IS are:
\begin{itemize}
\item In case of success, returns:
\begin{quote}
\dcsProtoMsg{\dcsProtoKeyword{OK}<b>PhyMachList}
\end{quote}
where \dcsProtoField{PhyMachList} is a list of entry messages:
\begin{quote}
\dcsProtoMsg{PHY\_ID<b>PHY\_IP<b>MM\_PORT}
\end{quote}
where:
\begin{itemize}
\item \dcsProtoField{PHY\_IP}: string containing the IP address of a physical machine.
\item \dcsProtoField{PHY\_ID}: integer number \dcsSSSecRef{formats-intnum} representing the identifier of a physical machine.
\item \dcsProtoField{MM\_PORT}: integer number \dcsSSSecRef{formats-intnum} representing the TCP port of the MM.
\end{itemize}
terminated by a dot message:
\begin{quote}
\dcsProtoMsg{\dcsProtoKeyword{$\ldotp$}}
\end{quote}
indicating the end of the list.
In case of empty list the following message is returned:
\begin{quote}
\dcsProtoMsg{\dcsProtoKeyword{OK}<b>$\ldotp$}
\end{quote}
\item \dcsProtoMsg{\dcsProtoKeyword{ERR}<b>CODE} otherwise, where \dcsProtoField{CODE} is an integer number representing an error code.
\end{itemize}

\subsubsection{LISTPHYMACHSTATUS -- \emph{List of Physical Machines along with Resource Utilization} request}

Sent by a SC (or other clients) to the IS when it wants to retrieve the list of all registered physical machines along with the status of their resources utilization.
\begin{quote}
\dcsProtoMsg{\dcsProtoKeyword{LISTPHYMACHSTATUS}}
\end{quote}
Possible replies from the IS are:
\begin{itemize}
\item In case of success, returns:
\begin{quote}
\dcsProtoMsg{\dcsProtoKeyword{OK}<b>PhyMachList}
\end{quote}
where \dcsProtoField{PhyMachList} is a list of entry messages:
\begin{quote}
\dcsProtoMsg{PHY\_ID<b>AVAIL\_CPU<b>AVAIL\_RAM<b>\\AVAIL\_DISK<b>NETSPEED}
\end{quote}
terminated by a dot message:
\begin{quote}
\dcsProtoMsg{\dcsProtoKeyword{$\ldotp$}}
\end{quote}
indicating the end of the list.

The fields in each entry has the following meaning:
\begin{itemize}
\item \dcsProtoField{PHY\_ID}: integer number \dcsSSSecRef{formats-intnum} representing the identifier of a physical machine.
\item \dcsProtoField{AVAIL\_CPU}: real number \dcsSSSecRef{formats-realnum} representing the available number of processors expressed as a fraction of the total number of CPU/Core processors:
{\small
\begin{multline*}
\dcsFuncName{NumOfCpus}(\dcsMathId{PHY\_ID}) \\- \sum_{\begin{smallmatrix}\dcsMathId{VM\_ID} \\ \text{on \dcsMathId{PHY\_ID}}\end{smallmatrix}} \dcsFuncName{AllocCpuFrac}(\dcsMathId{VM\_ID})
\end{multline*}
}
Admissibile values are in the range of $[0,\dcsFuncName{NumOfCpus(PHY\_ID)}]$.
\item \dcsProtoField{AVAIL\_RAM}: real number \dcsSSSecRef{formats-realnum} representing the available RAM expressed as a fraction of the total RAM size:
{\small
\begin{multline*}
1 - \sum_{\begin{smallmatrix}\dcsMathId{VM\_ID} \\ \text{on } \dcsMathId{PHY\_ID}\end{smallmatrix}} \dcsFuncName{AllocRamFrac}(\dcsMathId{VM\_ID})
\end{multline*}
}
Admissibile values are in the range of $[0,1]$.
\item \dcsProtoField{AVAIL\_DISK}: real number \dcsSSSecRef{formats-realnum} representing the available disk expressed as a fraction of the total disk size:
{\small
\begin{multline*}
1 - \sum_{\begin{smallmatrix}\dcsMathId{VM\_ID} \\ \text{on \dcsMathId{PHY\_ID}}\end{smallmatrix}} \dcsFuncName{AllocDiskFrac}(\dcsMathId{VM\_ID})
\end{multline*}
}
Admissibile values are in the range of $[0,1]$.
\item \dcsProtoField{NETSPEED}: string representing the total speed of the network interface card, expressed as an integer number followed by a unit of measurement specifier \dcsSSSecRef{formats-net}.
\end{itemize}
In case of empty list the following message is returned:
\begin{quote}
\dcsProtoMsg{\dcsProtoKeyword{OK}<b>$\ldotp$}
\end{quote}
\item \dcsProtoMsg{\dcsProtoKeyword{ERR}<b>CODE} otherwise, where \dcsProtoField{CODE} is an integer number representing an error code.
\end{itemize}

\subsubsection{LISTREPO -- \emph{List of Repositories} request}

Sent by a TAAROA component to the IS when it wants to know the list of available RMs.
\begin{quote}
\dcsProtoMsg{\dcsProtoKeyword{LISTREPO}}
\end{quote}
Possible replies from the IS are:
\begin{itemize}
\item In case of success, returns:
\begin{quote}
\dcsProtoMsg{\dcsProtoKeyword{OK}<b>RepoList}
\end{quote}
where \dcsProtoField{RepoList} is a list of entry messages:
\begin{quote}
\dcsProtoMsg{REPO\_ID<b>IP\_ADDR<b>PORT<b>\\\dcsProtoFieldEnc{USER\_NAME}<b>\dcsProtoFieldEnc{PASSWD}}
\end{quote}
where:
\begin{itemize}
\item \dcsProtoField{IP\_ADDR}: string containing the IP address of the repository manager service.
\item \dcsProtoField{PORT}: integer number \dcsSSSecRef{formats-intnum} representing the TCP port on which the repository manager service must be contacted.
\item \dcsProtoField{USER\_NAME}: string containing the username that must be used to authenticate with the RM.
\item \dcsProtoField{PASSWD}: string containing the password that must be used to authenticate with the RM.
\end{itemize}
terminated by a dot message:
\begin{quote}
\dcsProtoMsg{\dcsProtoKeyword{$\ldotp$}}
\end{quote}
indicating the end of the list.
In case of empty list the following message is returned:
\begin{quote}
\dcsProtoMsg{\dcsProtoKeyword{OK}<b>$\ldotp$}
\end{quote}
\item \dcsProtoMsg{\dcsProtoKeyword{ERR}<b>CODE} otherwise, where \dcsProtoField{CODE} is an integer number representing an error code.
\end{itemize}

\subsubsection{LISTSERV -- \emph{List of Services} request}

Sent by a TC to the IS when it wants to retrieve the list of all registered services.
\begin{quote}
\dcsProtoMsg{\dcsProtoKeyword{LISTSERV}}
\end{quote}
Possible replies from the IS are:
\begin{itemize}
\item In case of success, returns:
\begin{quote}
\dcsProtoMsg{\dcsProtoKeyword{OK}<b>ServList}
\end{quote}
where \dcsProtoField{ServList} is a list of entry messages:
\begin{quote}
\dcsProtoMsg{S\_ID<b>\dcsProtoFieldEnc{NAME}<b>RM\_ID<b>\\RM\_IP<b>RM\_PORT}
\end{quote}
where:
\begin{itemize}
\item \dcsProtoField{S\_ID}: integer number \dcsSSSecRef{formats-intnum} representing the service identifier.
\item \dcsProtoField{NAME}: string representing the symbolic name of the service.
\item \dcsProtoField{RM\_ID}: integer number \dcsSSSecRef{formats-intnum} representing the RM identifier.
\item \dcsProtoField{RM\_IP}: string representing the IP address of the RM.
\item \dcsProtoField{RM\_PORT}: integer number \dcsSSSecRef{formats-intnum} representing the TCP port of the RM.
\end{itemize}
terminated by a dot message:
\begin{quote}
\dcsProtoMsg{\dcsProtoKeyword{$\ldotp$}}
\end{quote}
indicating the end of the list.
In case of empty list the following message is returned:
\begin{quote}
\dcsProtoMsg{\dcsProtoKeyword{OK}<b>$\ldotp$}
\end{quote}
\item \dcsProtoMsg{\dcsProtoKeyword{ERR}<b>CODE} otherwise, where \dcsProtoField{CODE} is an integer number representing an error code.
\end{itemize}

\subsubsection{LISTVM -- \emph{List of Virtual Machines from Service} request}

Sent by a TC to the IS when it wants to retrieve the list of all submitted services (virtual machines) for the given service \dcsProtoField{S\_ID}.
\begin{quote}
\dcsProtoMsg{\dcsProtoKeyword{LISTVM}<b>S\_ID}
\end{quote}
where:
\begin{itemize}
\item \dcsProtoField{S\_ID}: integer number \dcsSSSecRef{formats-intnum} representing the service identifier.
\end{itemize}
Possible replies from the IS are:
\begin{itemize}
\item In case of success, returns:
\begin{quote}
\dcsProtoMsg{\dcsProtoKeyword{OK}<b>VMList}
\end{quote}
where \dcsProtoField{VMList} is a list of entry messages:
\begin{quote}
\dcsProtoMsg{VM\_ID<b>PHY\_ID<b>VM\_LOCAL\_ID<b>\\VIRT\_IP<b>STATUS}
\end{quote}
where:
\begin{itemize}
\item \dcsProtoField{VM\_ID}: integer number \dcsSSSecRef{formats-intnum} containing the submitted service (virtual machine) identifier.
\item \dcsProtoField{PHY\_ID}: integer number \dcsSSSecRef{formats-intnum} containing the identifier of the physical machine.
\item \dcsProtoField{VM\_LOCAL\_ID}: string containing the identifier used by the MM to uniquely retrieve a VM.
\item \dcsProtoField{VIRT\_IP}: string containing the IP address of the virtual machine on which the service is running.
\item \dcsProtoField{STATUS}: integer number \dcsSSSecRef{formats-intnum} representing the execution status of the submitted service (\dcsSSSecRef{concepts-vm}).
\end{itemize}
terminated by a dot message:
\begin{quote}
\dcsProtoMsg{\dcsProtoKeyword{$\ldotp$}}
\end{quote}
indicating the end of the list.
In case of empty list the following message is returned:
\begin{quote}
\dcsProtoMsg{\dcsProtoKeyword{OK}<b>$\ldotp$}
\end{quote}
\item \dcsProtoMsg{\dcsProtoKeyword{ERR}<b>CODE} otherwise, where \dcsProtoField{CODE} is an integer number representing an error code.
\end{itemize}

\subsubsection{REGPHYMACH -- \emph{Physical Machine Registration} request}

Sent by a MM to the IS for registering a specific physical machine.
\begin{quote}
\dcsProtoMsg{\dcsProtoKeyword{REGPHYMACH}<b>PHY\_IP<b>\dcsProtoFieldEnc{CPUTYPE}<b>\\NCPU<b>CPUCLOCK<b>RAMSIZE<b>\\DISKSIZE<b>NETSPEED<b>MAX\_VM\_NUMBER<b>\\\dcsProtoFieldEnc{MACH\_USERNAME}<b>\dcsProtoFieldEnc{MACH\_PASSWORD}<b>\\\dcsProtoFieldEnc{XM\_USERNAME}<b>\dcsProtoFieldEnc{XM\_PASSWORD}<b>\\MM\_PORT}
\end{quote}
where:
\begin{itemize}
\item \dcsProtoField{PHY\_IP}: string representing the physical machine IP address.
\item \dcsProtoField{CPUTYPE}: string representing the model or architecture or type of the CPU installed on the machine.
\item \dcsProtoField{NCPU}: integer number \dcsSSSecRef{formats-intnum} representing the total number of CPU processors/cores installed on the machine.
\item \dcsProtoField{CPUCLOCK}: string representing the clock frequency of a single CPU processor/core of the machine.
See \dcsSSSecRef{formats-freq} for the specification of frequency unit of measurement.
If no unit specifier character is specified, the \emph{MegaHertz} unit of measurement is assumed as default.
\item \dcsProtoField{RAMSIZE}: string representing the total memory available on the machine.
See \dcsSSSecRef{formats-mem} for the specification of memory unit of measurement.
If no unit specifier character is specified, the \emph{Megabyte} unit is assumed as default.
\item \dcsProtoField{DISKSIZE}: string representing the total disk space available on the machine.
See \dcsSSSecRef{formats-mem} for the specification of memory unit of measurement.
If no unit specifier character is specified, the \emph{Megabyte} unit is assumed as default.
\item \dcsProtoField{NETSPEED}: string representing the speed of the (main) network card installed on the machine.
See \dcsSSSecRef{formats-net} for the specification of net speed unit of measurement.
If no unit specifier character is specified, the \emph{Mbit/s} unit is assumed as default.
\item \dcsProtoField{MAX\_VM\_NUMBER}: integer number \dcsSSSecRef{formats-intnum} representing the maximum allowed number of running virtual machines. The value $-1$ means ``no limit''.
\item \dcsProtoField{MACH\_USERNAME}: string representing the name of the user used for logging in the machine.
\item \dcsProtoField{MACH\_PASSWORD}: string representing the password of the user used for logging in the machine.
\item \dcsProtoField{XM\_USERNAME}: string representing the name of the user used for controlling the Xen Manager.
\item \dcsProtoField{XM\_PASSWORD}: string representing the password of the user used for controlling the Xen Manager.
\item \dcsProtoField{MM\_PORT}: integer number \dcsSSSecRef{formats-intnum} representing the TCP port number where the MM is accepting connections.
\end{itemize}
Possible replies from the IS are:
\begin{itemize}
\item \dcsProtoMsg{\dcsProtoKeyword{OK}<b>PHY\_ID} in case of success, where \dcsProtoField{PHY\_ID} is the integer identifier of the new registered physical machine.
\item \dcsProtoMsg{\dcsProtoKeyword{ERR}<b>CODE} otherwise, where \dcsProtoField{CODE} is an integer number representing an error code.
\end{itemize}

\subsubsection{REGREPO -- \emph{Repository Manager Registration} request}

Sent by a RM to the IS for registering itself.
\begin{quote}
\dcsProtoMsg{\dcsProtoKeyword{REGREPO}<b>IP\_ADDR<b>PORT<b>\\\dcsProtoFieldEnc{USER\_NAME}<b>\dcsProtoFieldEnc{PASSWD}}
\end{quote}
where:
\begin{itemize}
\item \dcsProtoField{IP\_ADDR}: string containing the IP address of the RM.
\item \dcsProtoField{PORT}: integer number \dcsSSSecRef{formats-intnum} containing the TCP port on which the RM waits for requests.
\item \dcsProtoField{USER\_NAME}: string containing the username that must be used to authenticate with the RM.
\item \dcsProtoField{PASSWD}: string containing the password that must be used to authenticate with the RM.
\end{itemize}
Possible replies from the IS are:
\begin{itemize}
\item \dcsProtoMsg{\dcsProtoKeyword{OK}<b>RM\_ID} in case of success, where \dcsProtoField{RM\_ID} is the integer identifier of the new registered RM.
\item \dcsProtoMsg{\dcsProtoKeyword{ERR}<b>CODE} otherwise, where \dcsProtoField{CODE} is an integer number representing an error code.
\end{itemize}

\subsubsection{REGSERV -- \emph{Service Registration} request}

Sent by a RM to the IS when it wants to register a new service.
\begin{quote}
\dcsProtoMsg{\dcsProtoKeyword{REGSERV}<b>RM\_ID<b>\dcsProtoFieldEnc{NAME}<b>\dcsProtoField{REQ\_DISK}}
\end{quote}
where:
\begin{itemize}
\item \dcsProtoField{RM\_ID}: integer number \dcsSSSecRef{formats-intnum} containing the RM identifier.
\item \dcsProtoField{NAME}: string containing the symbolic name of the service.
\item \dcsProtoField{REQ\_DISK}: string representing the disk requirements.
See \dcsSSSecRef{formats-mem} for the specification of disk unit of measurement.
If no unit specifier character is specified, the \emph{Kilobyte} unit is assumed as default.
\end{itemize}
Possible replies from the IS are:
\begin{itemize}
\item \dcsProtoMsg{\dcsProtoKeyword{OK}<b>S\_ID} in case of success, where \dcsProtoField{S\_ID} is the integer identifier of the new registered service.
\item \dcsProtoMsg{\dcsProtoKeyword{ERR}<b>CODE} otherwise, where \dcsProtoField{CODE} is an integer number representing an error code.
\end{itemize}

\subsubsection{REGVM -- \emph{Virtual Machine Registration} request}

Sent by a MM to the IS when it wants to register a running VM (i.e., a VM that has been started on a physical machine).
\begin{quote}
\dcsProtoMsg{\dcsProtoKeyword{REGVM}<b>S\_ID<b>PHY\_ID<b>VM\_LOCAL\_ID<b>\\VIRT\_IP<b>ALLOCATED\_CPU<b>\\ALLOCATED\_RAM<b>ALLOCATED\_DISK}
\end{quote}
where:
\begin{itemize}
\item \dcsProtoField{S\_ID}: integer number \dcsSSSecRef{formats-intnum} containing the service identifier.
\item \dcsProtoField{PHY\_ID}: integer number \dcsSSSecRef{formats-intnum} containing the identifier of the physical machine.
\item \dcsProtoField{VM\_LOCAL\_ID}: string containing an identifier used by the MM to uniquely retrieve a VM.
\item \dcsProtoField{VIRT\_IP}: string containing the IP address of the physical machine on which the VM is running.
\item \dcsProtoField{ALLOCATED\_CPU}: real number \dcsSSSecRef{formats-realnum} representing the number of CPU/Core processors allocated to the VM.
\item \dcsProtoField{ALLOCATED\_RAM}: real number \dcsSSSecRef{formats-realnum} representing the amount of RAM allocated to the VM.
\item \dcsProtoField{ALLOCATED\_DISK}: real number \dcsSSSecRef{formats-realnum} representing the amount of disk allocated to the VM.
\end{itemize}
Possible replies from the IS are:
\begin{itemize}
\item \dcsProtoMsg{\dcsProtoKeyword{OK}<b>VM\_ID} in case of success, where \dcsProtoField{VM\_ID} is the integer identifier of the new registered virtual machine.
\item \dcsProtoMsg{\dcsProtoKeyword{ERR}<b>CODE} otherwise, where \dcsProtoField{CODE} is an integer number representing an error code.
\end{itemize}

\subsubsection{SRVPROTOVER -- \emph{Protocol Version} request}

Sent by a client to the IS for getting information about the TAAROA protocol version implemented by the IS server.
\begin{quote}
\dcsProtoMsg{\dcsProtoKeyword{SRVPROTOVER}}
\end{quote}
Possible replies from the IS are:
\begin{itemize}
\item In case of success:
\begin{quote}
\dcsProtoMsg{\dcsProtoKeyword{OK}<b>VERSION}
\end{quote}
where:
\begin{itemize}
\item \dcsProtoField{VERSION}: string containing the TAAROA protocol version implemented by the server.
\end{itemize}
\item \dcsProtoMsg{\dcsProtoKeyword{ERR}<b>CODE} otherwise, where \dcsProtoField{CODE} is an integer number representing an error code.
\end{itemize}

\subsubsection{UNREGPHYMACH -- \emph{Physical Machine Unregistration} request}

Sent by a MM (or other clients) to the IS for unregistering a specific physical machine.
\begin{quote}
\dcsProtoMsg{\dcsProtoKeyword{UNREGPHYMACH}<b>PHY\_ID}
\end{quote}
where:
\begin{itemize}
\item \dcsProtoField{PHY\_ID}: integer number \dcsSSSecRef{formats-intnum} representing the physical machine identifier.
\end{itemize}
Possible replies from the IS are:
\begin{itemize}
\item \dcsProtoMsg{\dcsProtoKeyword{OK}<b>PHY\_ID} in case of success, where \dcsProtoField{PHY\_ID} is the integer identifier of the unregistered physical machine (the same received in the request message).
\item \dcsProtoMsg{\dcsProtoKeyword{ERR}<b>CODE} otherwise, where \dcsProtoField{CODE} is an integer number representing an error code.
\end{itemize}
\emph{Side Effects}: all virtual machines running on this machine MUST be unregistered as well.

\subsubsection{UNREGREPO -- \emph{Repository Manager Unregistration} request}

Sent by a RM (or other clients) to the IS for unregistering itself (a specific repository manager).
\begin{quote}
\dcsProtoMsg{\dcsProtoKeyword{UNREGREPO}<b>RM\_ID}
\end{quote}
where:
\begin{itemize}
\item \dcsProtoField{RM\_ID}: integer number \dcsSSSecRef{formats-intnum} representing the repository manager identifier.
\end{itemize}
Possible replies from the IS are:
\begin{itemize}
\item \dcsProtoMsg{\dcsProtoKeyword{OK}<b>RM\_ID} in case of success, where \dcsProtoField{RM\_ID} is the integer identifier of the unregistered repository manager (the same received in the request message).
\item \dcsProtoMsg{\dcsProtoKeyword{ERR}<b>CODE} otherwise, where \dcsProtoField{CODE} is an integer number representing an error code.
\end{itemize}
\emph{Side Effects}: all services and related virtual machines registered by this RM MUST be unregistered as well.

\subsubsection{UNREGSERV -- \emph{Service Unregistration} request}

Sent by a RM to the IS for unregistering a specific service.
\begin{quote}
\dcsProtoMsg{\dcsProtoKeyword{UNREGSERV}<b>S\_ID}
\end{quote}
where:
\begin{itemize}
\item \dcsProtoField{S\_ID}: integer number \dcsSSSecRef{formats-intnum} representing the service identifier.
\end{itemize}
Possible replies from the IS are:
\begin{itemize}
\item \dcsProtoMsg{\dcsProtoKeyword{OK}<b>S\_ID} in case of success, where \dcsProtoField{S\_ID} is the integer identifier of the unregistered service (the same received in the request message).
\item \dcsProtoMsg{\dcsProtoKeyword{ERR}<b>CODE} otherwise, where \dcsProtoField{CODE} is an integer number representing an error code.
\end{itemize}
\emph{Side Effects}: all virtual machines associated to this service MUST be unregistered as well.

\subsubsection{UNREGVM -- \emph{Virtual Machine Unregistration} request}

Sent by a MM to the IS for unregistering a specific running service (virtual machine).
\begin{quote}
\dcsProtoMsg{\dcsProtoKeyword{UNREGVM}<b>VM\_ID}
\end{quote}
where:
\begin{itemize}
\item \dcsProtoField{VM\_ID}: integer number \dcsSSSecRef{formats-intnum} representing the virtual machine identifier.
\end{itemize}
Possible replies from the IS are:
\begin{itemize}
\item \dcsProtoMsg{\dcsProtoKeyword{OK}<b>VM\_ID} in case of success, where \dcsProtoField{VM\_ID} is the integer identifier of the unregistered virtual machine (the same received in the request message).
\item \dcsProtoMsg{\dcsProtoKeyword{ERR}<b>CODE} otherwise, where \dcsProtoField{CODE} is an integer number representing an error code.
\end{itemize}

%

\subsubsection{UPDATEVMSTATUS -- \emph{Virtual Machine Status Update} request}

Sent by a RM to the IS when it wants to update the execution status of a given submitted service (virtual machine).
\begin{quote}
\dcsProtoMsg{\dcsProtoKeyword{UPDATEVMSTATUS}<b>VM\_ID<b>STATUS}
\end{quote}
where:
\begin{itemize}
\item \dcsProtoField{VM\_ID}: integer number \dcsSSSecRef{formats-intnum} containing the submitted service (virtual machine) identifier.
\item \dcsProtoField{STATUS}: integer number \dcsSSSecRef{formats-intnum} representing the execution status of the submitted service (\dcsSSSecRef{concepts-vm}).
\end{itemize}
Possible replies from the IS are:
\begin{itemize}
	\item \dcsProtoMsg{\dcsProtoKeyword{OK}<b>STATUS} in case of success, where \dcsProtoField{STATUS} is the new execution status of the submitted service (virtual machine).
\item \dcsProtoMsg{\dcsProtoKeyword{ERR}<b>CODE} otherwise, where \dcsProtoField{CODE} is an integer number representing an error code.
\end{itemize}



\subsection{Messages issued to the Repository Manager} \label{ssec:rm}

\subsubsection{SRVPROTOVER -- \emph{Protocol Version} request}

Sent by a client to the RM for getting information about the TAAROA protocol version implemented by the RM server.
\begin{quote}
\dcsProtoMsg{\dcsProtoKeyword{SRVPROTOVER}}
\end{quote}
Possible replies from the RM are:
\begin{itemize}
\item In case of success:
\begin{quote}
\dcsProtoMsg{\dcsProtoKeyword{OK}<b>VERSION}
\end{quote}
where:
\begin{itemize}
\item \dcsProtoField{VERSION}: string containing the TAAROA protocol version implemented by the server.
\end{itemize}
\item \dcsProtoMsg{\dcsProtoKeyword{ERR}<b>CODE} otherwise, where \dcsProtoField{CODE} is an integer number representing an error code.
\end{itemize}

\subsubsection{STOPVM -- \emph{Service Stop} request}

Sent by a SC to the RM for stopping a given submitted service (virtual machine).
\begin{quote}
\dcsProtoMsg{\dcsProtoKeyword{STOPVM}<b>VM\_ID}
\end{quote}
where:
\begin{itemize}
\item \dcsProtoField{VM\_ID}: integer number \dcsSSSecRef{formats-intnum} representing the submitted service (virtual machine) identifier.
\end{itemize}
Possible replies from the RM are:
\begin{itemize}
\item \dcsProtoMsg{\dcsProtoKeyword{OK}<b>VM\_ID} in case of success, where \dcsProtoField{VM\_ID} is the identifier of the stopped submitted service (virtual machine).
\item \dcsProtoMsg{\dcsProtoKeyword{ERR}<b>CODE} otherwise, where \dcsProtoField{CODE} is an integer number representing an error code.
\end{itemize}

\subsubsection{SUBMITVM -- \emph{Service Submission} request}

Sent by a SC to the RM for submitting a given service.
\begin{quote}
\dcsProtoMsg{\dcsProtoKeyword{SUBMITVM}<b>S\_ID<b>PHY\_ID}
\end{quote}
where:
\begin{itemize}
\item \dcsProtoField{S\_ID}: integer number \dcsSSSecRef{formats-intnum} representing the service identifier.
\item \dcsProtoField{PHY\_ID}: integer number \dcsSSSecRef{formats-intnum} representing the identifier of the physical machine where the service has to be executed.
\end{itemize}
Possible replies from the RM are:
\begin{itemize}
\item \dcsProtoMsg{\dcsProtoKeyword{OK}<b>VM\_ID} in case of success, where \dcsProtoField{VM\_ID} is the integer identifier of the virtual machine where the submitted service (virtual machine) is running.
\item \dcsProtoMsg{\dcsProtoKeyword{ERR}<b>CODE} otherwise, where \dcsProtoField{CODE} is an integer number representing an error code.
\end{itemize}



\subsection{Messages issued to the Scheduler} \label{ssec:sc}

\subsubsection{SRVPROTOVER -- \emph{Protocol Version} request}

Sent by a client to the SC for getting information about the TAAROA protocol version implemented by the SC server.
\begin{quote}
\dcsProtoMsg{\dcsProtoKeyword{SRVPROTOVER}}
\end{quote}
Possible replies from the SC are:
\begin{itemize}
\item In case of success:
\begin{quote}
\dcsProtoMsg{\dcsProtoKeyword{OK}<b>VERSION}
\end{quote}
where:
\begin{itemize}
\item \dcsProtoField{VERSION}: string containing the TAAROA protocol version implemented by the server.
\end{itemize}
\item \dcsProtoMsg{\dcsProtoKeyword{ERR}<b>CODE} otherwise, where \dcsProtoField{CODE} is an integer number representing an error code.
\end{itemize}

\subsubsection{STOPSERV -- \emph{Service Stop} request}

Sent by a TC to the SC for stopping a given submitted service (virtual machine).
\begin{quote}
\dcsProtoMsg{\dcsProtoKeyword{STOPSERV}<b>VM\_ID}
\end{quote}
where:
\begin{itemize}
\item \dcsProtoField{VM\_ID}: integer number \dcsSSSecRef{formats-intnum} representing the identifier of a running service instance.
\end{itemize}
Possible replies from the SC are:
\begin{itemize}
\item \dcsProtoMsg{\dcsProtoKeyword{OK}<b>VM\_ID} in case of success, where \dcsProtoField{VM\_ID} is the identifier of the stopped service instance.
\item \dcsProtoMsg{\dcsProtoKeyword{ERR}<b>CODE} otherwise, where \dcsProtoField{CODE} is an integer number representing an error code.
\end{itemize}

\subsubsection{SUBMITSERV -- \emph{Service Submission} request}

Sent by a TC to the SC for starting the execution of a given service.
\begin{quote}
\dcsProtoMsg{\dcsProtoKeyword{SUBMITSERV}<b>S\_ID}
\end{quote}
where:
\begin{itemize}
\item \dcsProtoField{S\_ID}: integer number \dcsSSSecRef{formats-intnum} representing the service identifier.
\end{itemize}
Possible replies from the SC are:
\begin{itemize}
\item \dcsProtoMsg{\dcsProtoKeyword{OK}<b>VM\_ID} in case of success, where \dcsProtoField{VM\_ID} is the integer identifier of the running instance of the given service.
\item \dcsProtoMsg{\dcsProtoKeyword{ERR}<b>CODE} otherwise, where \dcsProtoField{CODE} is an integer number representing an error code.
\end{itemize}



\subsection{Messages issued to the Machine Manager} \label{ssec:mm}

\subsubsection{SRVPROTOVER -- \emph{Protocol Version} request}

Sent by a client to the MM for getting information about the TAAROA protocol version implemented by the MM server.
\begin{quote}
\dcsProtoMsg{\dcsProtoKeyword{SRVPROTOVER}}
\end{quote}
Possible replies from the MM are:
\begin{itemize}
\item In case of success:
\begin{quote}
\dcsProtoMsg{\dcsProtoKeyword{OK}<b>VERSION}
\end{quote}
where:
\begin{itemize}
\item \dcsProtoField{VERSION}: string containing the TAAROA protocol version implemented by the server.
\end{itemize}
\item \dcsProtoMsg{\dcsProtoKeyword{ERR}<b>CODE} otherwise, where \dcsProtoField{CODE} is an integer number representing an error code.
\end{itemize}

\subsubsection{STARTVM -- \emph{Virtual Machine Execution} request}

Sent by a RM to the MM for starting a virtual machine given all the file necessary for running it.
\begin{quote}
\dcsProtoMsg{\dcsProtoKeyword{STARTVM}<b>S\_ID + <VM\_IMAGE>}
\end{quote}
where:
\begin{itemize}
\item \dcsProtoField{S\_ID}: integer number \dcsSSSecRef{formats-intnum} representing the service identifier.
\item \dcsProtoField{<VM\_IMAGE>}: all the file necessary for running the virtual machine.
\end{itemize}
Possible replies from the MM are:
\begin{itemize}
\item \dcsProtoMsg{\dcsProtoKeyword{OK}<b>VM\_ID} in case of success, where \dcsProtoField{VM\_ID} is the integer identifier of the virtual machine where the submitted service (virtual machine) is running.
\item \dcsProtoMsg{\dcsProtoKeyword{ERR}<b>CODE} otherwise, where \dcsProtoField{CODE} is an integer number representing an error code.
\end{itemize}

\subsubsection{STOPVM -- \emph{Virtual Machine Stop} request}

Sent by a RM to the MM for stopping a given submitted service (virtual machine).
\begin{quote}
\dcsProtoMsg{\dcsProtoKeyword{STOPVM}<b>VM\_LOCAL\_ID}
\end{quote}
where:
\begin{itemize}
\item \dcsProtoField{VM\_LOCAL\_ID}: integer number \dcsSSSecRef{formats-intnum} representing the submitted service (virtual machine) identifier.
\end{itemize}
Possible replies from the RM are:
\begin{itemize}
\item \dcsProtoMsg{\dcsProtoKeyword{OK}<b>0} in case of success.
\item \dcsProtoMsg{\dcsProtoKeyword{ERR}<b>CODE} otherwise, where \dcsProtoField{CODE} is an integer number representing an error code.
\end{itemize}



\subsection{Workflows Diagrams} \label{ssec:protocol-workflow}

To illustrate the communication protocol at work, we present in this section the two sample workflows previously described in \dcsSSecRef{workflow}.
The type of modelling diagram used for introducing the interaction between the different TAAROA component is the UML \emph{sequence diagram} \cite{OMG2007UML}.
This kind of diagram shows how components communicate with each other in terms of a sequence of messages.
Furthermore, it indicates the lifespans of components relative to those messages.

\begin{figure*}
\centering
\centerline{\includegraphics[scale=0.50]{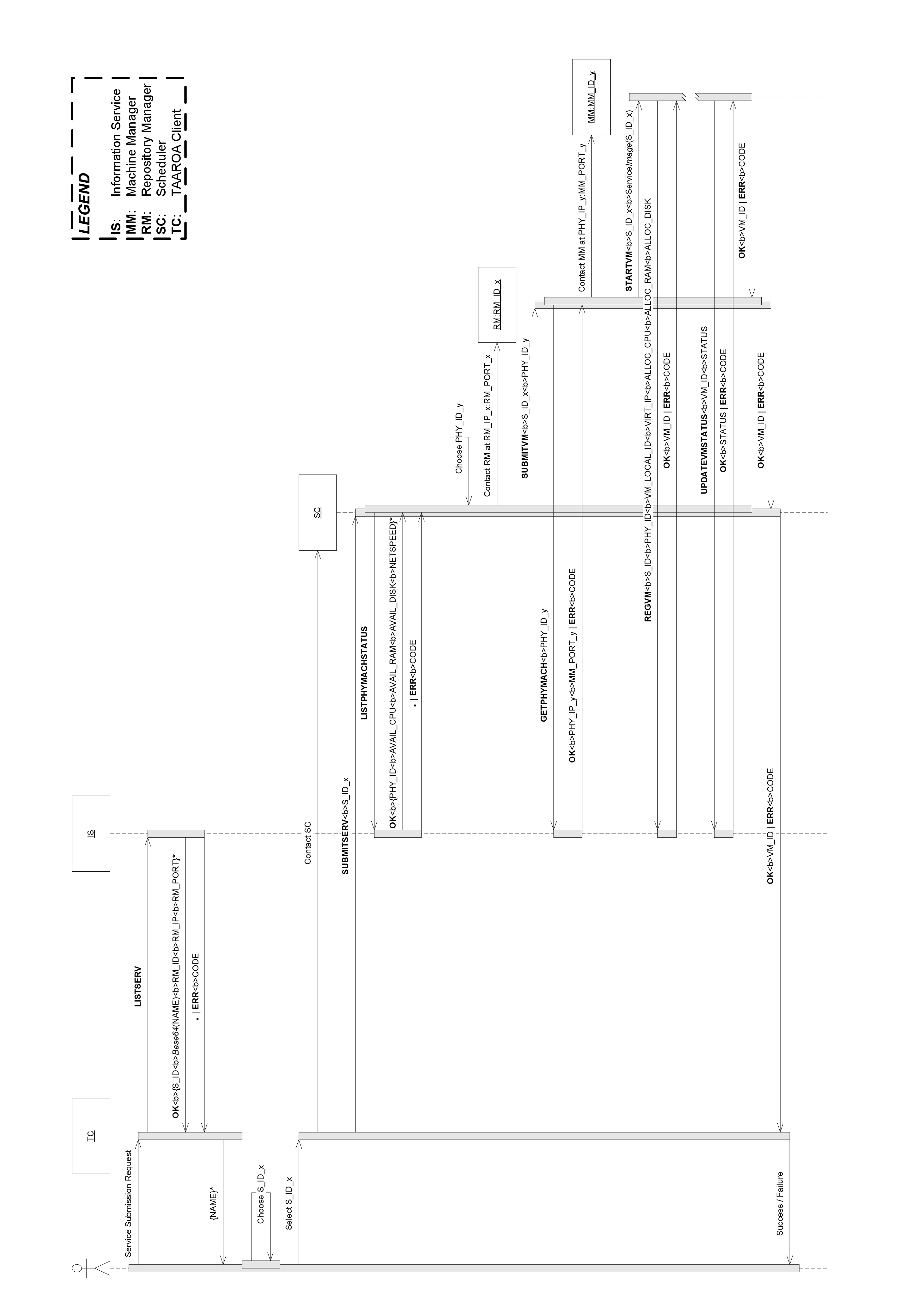}}
\caption{The sequence diagram for TAAROA service submission.}
\label{fig:workflow-svcstart-seq}
\end{figure*}

\dcsFigRef{workflow-svcstart-seq} shows the sequence diagram for the service submission request corresponding to the Service submission workflow presented in \dcsSSecRef{workflow}.

\begin{figure*}
\centering
\includegraphics[scale=0.50]{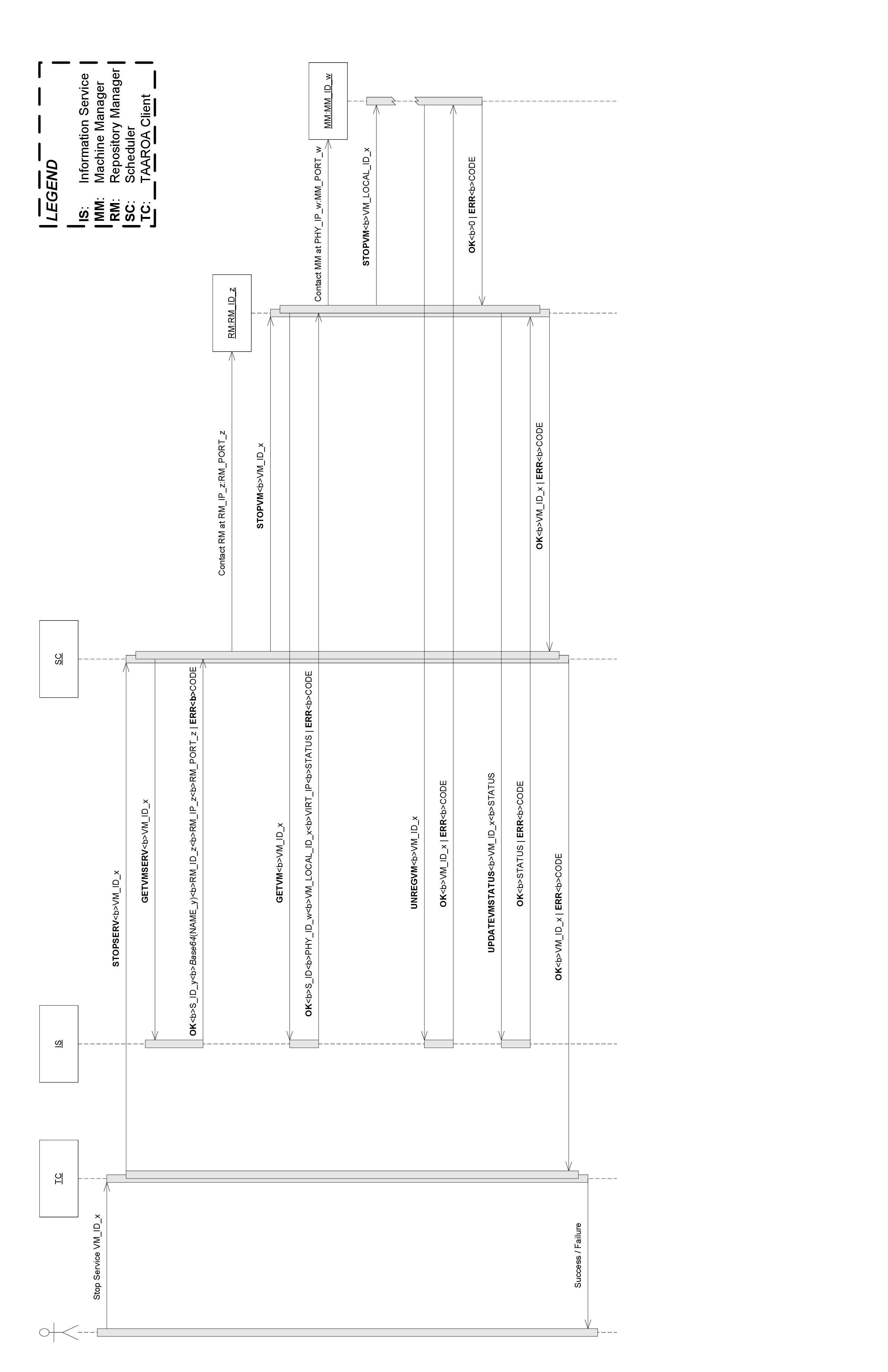}
\caption{The sequence diagram for TAAROA service stopping.}
\label{fig:workflow-svcstop-seq}
\end{figure*}

\dcsFigRef{workflow-svcstop-seq} shows the sequence diagram for the service stopping request related to the Service stopping workflow described in \dcsSSecRef{workflow}.




\section{Conclusions and Future Work} \label{sec:conclusion}

In this paper we presented the TAAROA middleware, a software system that tries to add the concept of service and Service Level Agreement (SLA) to the Grid computing paradigm, by using the virtualization technology.
The current version of TAAROA has some limitations.
The most important of these are the lack of a logic for mapping a high-level SLA specification to a low-level resource allocation and, as a consequence, the absence of a scheduling heuristic that properly assigns a Physical Machine to a Virtual Machine taking into account the preservation of SLA constraints.
This means that actually TAAROA is only able to provide a best-effort service: each service is scheduled for execution with a First-Come-First-Served policy and is assigned to the first available Physical Machine.
In the future, we plan to provide a better support for proactively or reactively avoiding SLA violations, by creating specific performance models and exploiting, for instance, the Virtual Machines migration.




\bibliography{paper}


\end{document}